\documentstyle [twocolumn,epsf]{mn}
\newcommand{\mincir}{\raise
-3.truept\hbox{\rlap{\hbox{$\sim$}}\raise4.truept\hbox{$<$}\ }}
\newcommand{\magcir}{\raise
-3.truept\hbox{\rlap{\hbox{$\sim$}}\raise4.truept\hbox{$>$}\ }}
\newcommand{\minmag}{\raise
-3.truept\hbox{\rlap{\hbox{$<$}}\raise5.truept\hbox{$<$}\ }}
\newcommand{\be}{\begin{equation}}
\newcommand{\ee}{\end{equation}}
 \newcommand{\ba}{\begin{eqnarray}}
\newcommand{\ea}{\end{eqnarray}}
\newcommand{\brr}{\begin{array}}
\newcommand{\err}{\end{array}}
\newcommand{\bc}{\begin{center}}
\newcommand{\ec}{\end{center}}

\newcommand{\omm}{{\Omega_{\rm m}}}
\newcommand{\epkk}{{E_{peak}}}
\newcommand{\tlag}{{\tau_{lag}}}
\newcommand{\trt}{{\tau_{RT}}}

\newcommand{\pbo}{{P_{\rm bolo}}}
\newcommand{\sbo}{{S_{\rm bolo}}}
\newcommand{\bd}{{$\ldots$}}
\newcommand{\cd}{{$\cdots$}}
\renewcommand{\baselinestretch}{1.0}

\title[Testing Gamma Ray Bursts as Standard Candles]
{Testing Gamma Ray Bursts as Standard Candles}
\author[S. Basilakos \& L. Perivolaropoulos]{Spyros Basilakos$^{1}$ \&
Leandros Perivolaropoulos$^{2}$.\\
\vspace{0.1cm}
$^1$ Academy of Athens Research Center for Astronomy \&
Applied Mathematics, Soranou Efessiou 4, 11-527 Athens, Greece \\
$^2$ Department of Physics, University of Ioannina, 451-10 Ioannina, Greece
}

\begin{document}

\maketitle

\begin{abstract}
Several interesting correlations among Gamma Ray Burst (GRB)
observables with available redshifts have been recently identified.
Proper evaluation and calibration of these correlations may
facilitate the use of GRBs as standard candles constraining the
expansion history of the universe up to redshifts of $z>6$. Here we
use the 69 GRB dataset recently compiled by Schaefer (2007)
and we test the calibration of five of the above
correlations ($1:\epkk-E_\gamma$, $2:\epkk-L$,  $3:\tlag-L$,
$4:V-L$, $5:\trt-L$) with respect to two potential sources of
systematics: Evolution with redshift and cosmological model used in
the calibration. In examining the model dependence we assume flat
$\Lambda$CDM and vary $\omm$. Our approach avoids the circularity problem
of previous studies since we do not fix $\omm$ to find the
correlation parameters. Instead we simultaneously minimize $\chi^2$
with respect to both the log-linear correlation parameters $a$, $b$
and the cosmological parameter $\omm$. We find no statistically
significant evidence for redshift dependence of $a$ and $b$ in any
of the correlation relations tested [the slopes of $a(z)$
and $b(z)$ are consistent with 0 at the $2\sigma$ level]. We also
find that one of the
five correlation relations tested ($\epkk-E_\gamma$) has a
significantly lower intrinsic dispersion compared to the other
correlations. For this correlation relation, the maximum likelihood
method ($\chi^2$ minimization) leads to $b_1=50.58\pm 0.04$,
$a_1=1.56\pm 0.11$, $\omm=0.30^{+1.60}_{-0.25}$ respectively. The
other four correlation relations minimize $\chi^2$ for a flat matter
dominated universe $\omm \simeq 1$. Finally, a cross-correlation
analysis between the GRBs and SnIa data for various values of $\omm$
has shown that the $E_{peak}-E_\gamma$ relation traces well the SnIa
regime (within $0.17\le z\le 1.755$). In particular, for
$\Omega_{\rm m}\simeq 0.15$ and $\Omega_{\rm m} \simeq 0.30$ we get the
highest correlation signal between the two populations. However, due
to the large error-bars in the cross-correlation analysis (small
number statistics) even the tightest correlation relation
($E_{peak}-E_\gamma$) provides much weaker constraints on $\omm$
than current SnIa data.

\end{abstract}

\begin{keywords}
cosmology:observations,distance scale; gamma rays:bursts
\end{keywords}

\vspace{1.0cm}

\section{Introduction}
Several cosmological observations
(Riess et al. 1998; Perlmutter et al. 1999; Tegmark et al. 2006;
Davis et al. 2007, Komatsu et al. 2008)
have converged during the last decade towards a cosmic expansion
history that involves a recent accelerating expansion of the
universe. This expansion has been attributed to an energy component
(dark energy) with negative pressure which dominates the universe at
late times and causes the observed accelerating expansion. The
simplest type of dark energy corresponds to the cosmological
constant (Padmanabhan 2003). Alternatively, modifications of
general relativity have also been utilized to explain the observed
cosmic acceleration (Boisseau et al. 2000; Perivolaropoulos 2005;
Caldwell, Cooray \& Melchiorri 2007;
Nesseris \& Perivolaropoulos 2006;
Jain \& Zhang 2007; Wang et al. 2007; Heavens, Kitching \& Verde 2007;
Nesseris \& Perivolaropoulos 2007).

The geometrical probes
(Lazkoz, Nesseris \& Perivolaropoulos 2008) used to map the cosmic
expansion history involve a combination of standard candles [Type Ia
supernovae (SnIa) Davis et al. 2007] and standard rulers [clusters,
CMB sound horizon detected through Baryon Acoustic Oscillations
(BAO; Percival et al. 2007 ) and through the CMB perturbations
angular power spectrum  Komatsu et al. 2008]. These observations
probe the integral of the Hubble expansion rate $H(z)$ either up to
redshifts of order $z\simeq 1-2$ (SnIa, BAO, clusters) or up to the
redshift of recombination ($z\simeq 1089$). Alternatively, dynamical
probes (Bertschinger 2006; Nesseris \& Perivolaropoulos 2008)
of the expansion history based on
measures of the growth rate of cosmological perturbations are also
confined to relatively low redshifts up to $z\simeq 1$. It is
therefore clear that the redshift range $\sim 2-1000$ is not
directly probed by any of the above observations. Even though many
models of dark energy predict a decelerating expansion in that
redshift range due to matter domination, the possibility of
non-trivial expansion properties at higher redshifts can not be
excluded. In order to investigate this possibility we need a visible
distance indicator at redshifts $z>2$.

Gamma Ray Bursts (GRBs) are the most violent and bright explosions
in the universe. They are produced by a highly relativistic bipolar
jet outflow from a compact source (Rhoads 1999; Piran 2004).
At present the most distant GRB (GRB 050904) is at a redshift
$z=6.3$ (Kawai et al. 2005). The fact that GRBs are detected up to
very high redshifts makes it tempting to try and use them as
standard candles that could be used to constrain the cosmological
expansion history in a similar way as SnIa. The problem is that GRBs
appear to be anything but standard candles: they have a very wide
range of isotropic equivalent luminosities and energy outputs.
Several suggestions have been made to calibrate them as better
standard candles by using correlations between various properties of
the prompt emission, and in some cases also the afterglow emission.
While there is good motivation for such cosmological applications of
GRBs, there are many practical difficulties.
Indeed, a serious problem that hampers such a straight forward
approach is the intrinsic
faintness of the nearby events which introduces a bias towards low (or high)
values of GRB observables and as a result of this, the extrapolation
of each of the correlations to low-z events is faced with
serious problems.
An additional problem is that
the GRB surveys suffer, due to the unknown flux limit,
from the well known degradation of sampling as
a function of redshift (Lloyd, Petrosian, \& Mallozzi 2000).
One might also expect a significant evolution of the
intrinsic properties of GRBs with redshift (also between
intermediate and high redshifts) which can be hard to disentangle
from cosmological effects. In addition, even after properly
accounting for the observed correlations, the scatter in the
luminosity of the standardized candles is still fairly large.
Finally, the calibration of the observed correlations require the
assumption of a cosmological model (luminosity distance vs redshift)
in order to convert the observed bolo-metric peak flux or bolo-metric
fluence to isotropic absolute luminosity $L$ or to a total
collimation corrected energy $E_{\gamma}$. The use of a cosmological
model to perform the calibration creates a circularity problem and a
model dependence of the obtained calibration.

Despite of the above difficulties, the potential benefits of
obtaining even approximate standard candles at redshifts as high as
$z>6$ has prompted a significant activity towards both testing the
usefulness of GRBs as standard candles
(Amati et al. 2002; Ghirlanda, Ghisellini \& Firmani 2006; Firmani et al. 2006;
Li 2007a; Li 2007b; Butler et al. 2007; Zhang \& Xie 2007;
Hooper \& Dodelson 2007)
and eagerly utilizing them to constrain cosmological parameters
(Schaefer et al. 2003; Zhang \& Meszaros 2004; Dai, Liang \&
Xu 2004; Di Girolamo, et al. 2005; Schafer 2007;
Bertolami \& Tavares Silva 2006; Wang \& Dai 2006;
Demianski, et al. (2006); Li et al. 2008a; Amati et al. 2008).
This activity has lead to a debate about the usefulness of GRBs as
standard candles with both discouraging
(Li 2007a; Li 2007b; Butler et al. 2007) and encouraging
(Amati et al. 2002; Ghirlanda et al. 2006; Firmani et al. 2006;
Schaefer 2007; Zhang \& Xie 2007; Hooper \& Dodelson 2007)
results.

The goal of the present study is to evaluate the current utility of
GRBs as standard candles and identify the correlation relations that
are more promising in the determination of the isotropic absolute
luminosity and total collimation corrected energy of the GRBs. The
evaluated correlation relations involve the relationship between a
measurable observable of the light curve (luminosity indicator) with
the GRB luminosity, given by the correlation relation in the form of
a power-law (Schaefer 2007), ie., $E_{\gamma} = B_{\gamma,
peak}\epkk^{a_{\gamma,peak}}$
(Ghirlanda, Ghisellini \& Lazzati 2004) \,
$L = B_{peak}\epkk^{a_{peak}}$,
(Schaefer et al. 2003) \,
 $L = B_{lag}
\tlag^{a_{lag}}$, (Norris, Marani \& Bonnell 2000)\,
$L=B_{v}V^{a_{v}}$,
(Fenimore \& Ramirez-Ruiz 2000) \,
and $L=B_{RT} \trt^{a_{RT}}$
(Schaefer 2007) [see section 2 for the definitions of the
above observable luminosity indicators]. These five correlation
relations can be denoted compactly as $R=B_j\; Q^{a_j}$ where $R$ is
proportional to the GRB absolute luminosity $L$, $Q$ is a GRB
observable, $j=1,..,5$ counts correlation relations while $B_j$,
$a_j$ are parameters to be calibrated using a $\chi^2$ minimization
with the GRB data.

We use the 69 GRB dataset compiled by Schaefer (2007)
and fit the logarithmic linear form of the
correlation relations ie., $\log R= b_j + a_j \log Q$ (where
$b_j=\log B_j$) using a maximum likelihood method of $\chi^2$
minimization (symmetric in $R$ and $Q$). We implement the following
tests:
\begin{itemize} \item We compare the quality of fit obtained with
each one of the above five correlation relations by evaluating
$\chi^2_{min}$ per degree of freedom in each case.
\item We split the dataset into four consecutive redshift bins and
evaluate the best fit parameters $a_j$, $b_j$ in each bin to test
for evolution effects of each correlation relation. \item The value
of $R$ (e.g. isotropic absolute luminosity) depends on both the
observed GRB bolo-metric peak flux and on the cosmological model
used to evaluate the luminosity distance (see eq. \ref{ldl}). We
thus evaluate the dependence of the quality of fit ($\chi^2_{min}$)
on the assumed cosmic expansion history $H(z)$. For concreteness we
assume a flat $\Lambda$CDM model and fit $a_j$, $b_j$ for various $\omm$.
This test is equivalent to simultaneously minimizing $\chi^2$ with
respect to both the calibration parameters $a_j$ and $b_j$ and the
cosmological parameter $\omm$. Such an approach avoids the
circularity problem discussed above and leads to an unambiguous
determination of the cosmological parameter $\omm$.
We would like to stress here that Schaefer (2007) claims
that during the $\omm$ fitting a marginalization over the
parameters $a_j$, $b_j$ could potentially solve the circularity problem.
However, if the marginalization is done over the $1\sigma$
range of the best fit parameter values then the information for the
imposed value of $\omm$ is still carried over leading to circularity.  If it
is done on a wider range then the errors on $\omm$ would be even larger than the ones obtained here.
\item Finally,
we perform a direct cross-correlation comparison of the GRB data
Schaefer (2007) with a recent SnIa dataset
Davis et al. (2007) in the redshift range of overlap.
\end{itemize}

The structure of this paper is the following: In the next section we
briefly review the five correlation relations studied and describe
our numerical fitting methods. In section 3 we present our
numerical results with respect to both the redshift dependence of
the correlation parameters and their dependence on the assumed
cosmological model. In section 4 we compare the GRBs with the SnIa
data in the redshift range of overlap. Finally, in section 5 we
conclude and discuss future prospects of this work.

\section{Correlation Relations - Method}
The correlation relations discussed in what follows connect GRB
observables with the isotropic absolute luminosity $L$ or the
collimation corrected energy $E_{\gamma}$ of the GRB. Such
observable properties of the GRBs include the peak energy, denoted
by $\epkk$, which is the photon energy at which the $\nu\,F_{\nu}$
spectrum is brightest; the jet opening angle, denoted by
$\theta_{\rm jet}$, which is the rest-frame time of the
achromatic break in the light curve of an afterglow; the time lag,
denoted by $\tlag$, which measures the time offset between high and
low energy GRB photons arriving on Earth and the variability,
denoted by $V$, which is the measurement of the ``spikiness'' or
``smoothness'' of the GRB light curve.
In the literature, there is a wide variety of choices for the
definition of $V$ (Fenimore \& Ramirez-Ruiz 2000; Reichart et al. 2001).
In this work, we follow the notations of Schaefer (2007)
in which the observed $V$ value varies as the inverse of the time
stretching, so the corresponding measured value must
be multiplied by $1+z$ to correct to the GRB rest frame.
An additional luminosity
indicator is the minimum rise time Schaefer (2007) denoted by
$\trt$, and taken to be the shortest time over which the light curve
rises by half the peak flux of the pulse.

These quantities appear to correlate with the GRB isotropic luminosity or
its total collimation-corrected energy. This property can not be
measured directly but rather it can be obtained through the
knowledge of either the bolo-metric peak flux, denoted by $\pbo$; or
bolo-metric fluence; denoted by $\sbo$, Schaefer (2007).
Therefore, the isotropic luminosity is given by:

\be
 L = 4\pi d^2_{L}(\Omega_{\rm m},z)P_{\rm bolo} \; \label{ldl}
\ee
and the total collimation-corrected energy reads:
\be  E_{\gamma}=4\pi
d^2_{L}(\Omega_{\rm m},z)S_{\rm bolo}F_{\rm beam}(1+z)^{-1} \; \label{egdl}
\ee where $F_{\rm beam}$ is the beaming factor\footnote{Note,
that $F_{beam}$ is calculated with the aid of $d_{L}(z)$
(see Sari, Piran \& Halpern 1999; Butler, Kocevski, \& Bloom 2008).
For small values of $\theta_{\rm jet}$, the dependence of $F_{beam}$ on $d_{L}(z)$
is found to be: $F_{beam} \propto d_{L}^{-1/2}$. Using the data from Schaefer 2007
($\Omega_{\rm m}=0.27$) we have to multiply the corresponding $F_{beam}$ value
by the following factor: $ [d_{L}(0.27,z)/d_{L}(\Omega_{\rm m},z)]^{1/2}$.}
($1-\cos{\theta_{\rm jet}}$).

\noindent The luminosity correlation relations are power-law
relations of either $L$ or $E_{\gamma}$ as a function of $\tlag$,
$V$,\, $E_{peak}$,\, $\trt$. Both $L$ and $E_{\gamma}$ depend
not only on the GRB observables $\pbo$ or $\sbo$, but also on the
cosmological parameters through the luminosity distance $d_L(z)$
which in a flat universe is expressed in terms of the Hubble
expansion rate $H(z)=H_0 E(z)$ as
\begin{eqnarray}
 \nonumber
 d_L(\omm,z) = (1+z)\frac{c}{H_0} \int_0^z \frac{dz'}{E(z')} \; ,
 \label{dlum1}
\end{eqnarray}
where $E^2(z) = \omm (1+z)^3 + \Omega_{\rm x}f_{\rm x}(z)$ and the
dimensionless dark energy density $f_{\rm x}(z)$ is given by
\be
 f_{\rm x}(z)=\frac{\rho_{\rm x}(z)}{\rho_{\rm x}(0)}=1
 \ee
where the last equality is valid in the case of $\Lambda$CDM
($\Omega_{\rm x}=\Omega_{\Lambda}$) which is assumed in what follows.

The relationship between a measurable observable of the light curve
(luminosity indicator) with the GRB luminosity is given by the
luminosity relation in the form of a power-law, i. e.,
$E_{\gamma} = B_{\gamma,
peak}\epkk^{a_{\gamma,peak}}$ (Ghirlanda et al. 2004)
\,$L = B_{peak}\epkk^{a_{peak}}$,
(Schaefer et al. 2003) \,
 $L = B_{lag}
\tlag^{a_{lag}}$, (Norris et al. 2000)\,
$L=B_{v}V^{a_{v}}$, (Fenimore \& Ramirez-Ruiz 2000) \,
and
$L=B_{RT} \trt^{a_{RT}}$
(Schaefer 2007).
The observed (on Earth) luminosity
indicators will have different values from those that would be
estimated in the rest frame of the GRB event. That is, the light
curves and spectra seen by the Earth-orbiting satellites suffer time
dilation and red-shifting. Therefore, the physical connection between
the indicators and the luminosity in the GRB rest frame must take
into account the observed indicators and correct them to the rest
frame of the GRB. For the temporal indicators, the observed
quantities must be divided by $1+z$ to correct the time dilation.
The observed $V$-value  must be multiplied by $1+z$ because it
varies inversely with time, and the observed $\epkk$ must be
multiplied by $1+z$ to correct the redshift dilation of the
spectrum. We have also rescaled the luminosity
indicators to dimensionless quantities by using the same values
employed by Schaefer (2007) in order to minimize correlations
between the normalization constant and the exponent during the
fitting, i. e., for the temporal luminosity we use 0.1 second, for
the variability 0.02, and for the energy indicator 300keV. For
example the effective GRB frame dimensionless value of $E_{peak}$
used in our analysis is $Q=\frac{E_{peak} (1+z)}{300keV}$ where
$E_{peak}$ is obtained from Table 1. The 69 GRB data of
Schaefer (2007) used in our analysis are shown in Table 1.

\begin{table*}
\tabcolsep 3pt
\begin{tabular}{|lllllllll|}
\hline
{\bf GRB} & z& $P_{\rm bolo}$ & $S_{\rm bolo}$ & $F_{\rm beam}$ & $\tau_{lag}$ &
       $V$ & $E_{\rm peak}$ &
       $\tau_{RT}$ \\
             &  &[${\rm erg/cm^{2}s}$] & [${\rm erg/cm^{2}}$]& &[sec]& &[keV] &[sec]  \\

\hline

970228\bd\bd&0.70 &7.3E-6 $\pm$ 4.3E-7 & \cd                 & \cd               &\cd           &0.0059$\pm$0.0008 &115$_{-38}^{+38}$      & 0.26 $\pm$ 0.04\\
970508\bd\bd&0.84 &3.3E-6 $\pm$ 3.3E-7 &8.09E-6 $\pm$ 8.1E-7 &0.0795$\pm$ 0.0204 &0.50$\pm$0.30 &0.0047$\pm$0.0009 &389$_{-[40]}^{+[40]}$  & 0.71 $\pm$ 0.06\\
970828\bd\bd&0.96 &1.0E-5 $\pm$ 1.1E-6 &1.23E-4 $\pm$ 1.2E-5 &0.0053$\pm$ 0.0014 &\cd           &0.0077$\pm$0.0007 &298$_{-[30]}^{+[30]}$  & 0.26 $\pm$ 0.07\\
971214\bd\bd&3.42 &7.5E-7 $\pm$ 2.4E-8 & \cd                 & \cd               &0.03$\pm$0.03 &0.0153$\pm$0.0006 &190$_{-[20]}^{+[20]}$  & 0.05 $\pm$ 0.02\\
980613\bd\bd&1.10 &3.0E-7 $\pm$ 8.3E-8 & \cd                 & \cd               &\cd           &\cd               &92 $_{-42  }^{+42  }$  & \cd            \\
980703\bd\bd&0.97 &1.2E-6 $\pm$ 3.6E-8 &2.83E-5 $\pm$ 2.9E-6 &0.0184$\pm$ 0.0027 &0.40$\pm$0.10 &0.0064$\pm$0.0003 &254$_{-[25]}^{+[25]}$  & 3.60 $\pm$ 0.5 \\
990123\bd\bd&1.61 &1.3E-5 $\pm$ 5.0E-7 &3.11E-4 $\pm$ 3.1E-5 &0.0024$\pm$ 0.0007 &0.16$\pm$0.03 &0.0175$\pm$0.0001 &604$_{-[60]}^{+[60]}$  & \cd            \\
990506\bd\bd&1.31 &1.1E-5 $\pm$ 1.5E-7 & \cd                 & \cd                 &0.04$\pm$0.02 &0.0131$\pm$0.0001 &283$_{-[30]}^{+[30]}$  & 0.17 $\pm$ 0.03\\
990510\bd\bd&1.62 &3.3E-6 $\pm$ 1.2E-7 &2.85E-5 $\pm$ 2.9E-6 &0.0021$\pm$ 0.0003 &0.03$\pm$0.01 &0.0100$\pm$0.0001 &126$_{-[10]}^{+[10]}$  & 0.14 $\pm$ 0.02\\
990705\bd\bd&0.84 &6.6E-6 $\pm$ 2.6E-7 &1.34E-4 $\pm$ 1.5E-5 &0.0035$\pm$ 0.0010 &\cd           &0.0210$\pm$0.0008 &189$_{-15  }^{+15  }$  & 0.05 $\pm$ 0.02\\
990712\bd\bd&0.43 &3.5E-6 $\pm$ 2.9E-7 &1.19E-5 $\pm$ 6.2E-7 &0.0136$\pm$ 0.0034 &\cd           &\cd               &65 $_{-10  }^{+10  }$  & \cd            \\
991208\bd\bd&0.71 &2.1E-5 $\pm$ 2.1E-6 & \cd                 & \cd               &\cd           &0.0037$\pm$0.0001 &190$_{-[20]}^{+[20]}$  & 0.32 $\pm$ 0.04\\
991216\bd\bd&1.02 &4.1E-5 $\pm$ 3.8E-7 &2.48E-4 $\pm$ 2.5E-5 &0.0030$\pm$ 0.0009 &0.03$\pm$0.01 &0.0130$\pm$0.0001 &318$_{-[30]}^{+[30]}$  & 0.08 $\pm$ 0.02\\
000131\bd\bd&4.50 &7.3E-7 $\pm$ 8.3E-8 & \cd                 & \cd               &\cd           &0.0053$\pm$0.0006 &163$_{-13  }^{+13  }$  & 0.12 $\pm$ 0.06\\
000210\bd\bd&0.85 &2.0E-5 $\pm$ 2.1E-6 & \cd                 & \cd               &\cd           &0.0041$\pm$0.0004 &408$_{-14  }^{+14  }$  & 0.38 $\pm$ 0.06\\
000911\bd\bd&1.06 &1.9E-5 $\pm$ 1.9E-6 & \cd                 & \cd               &\cd           &0.0235$\pm$0.0014 &986$_{-[100]}^{+[100]}$& 0.05 $\pm$ 0.02\\
000926\bd\bd&2.07 &2.9E-6 $\pm$ 2.9E-7 & \cd                 & \cd               &\cd           &0.0134$\pm$0.0013 &100$_{-7   }^{+7   }$  & 0.05 $\pm$ 0.03\\
010222\bd\bd&1.48 &2.3E-5 $\pm$ 7.2E-7 &2.45E-4 $\pm$ 9.1E-6 &0.0014$\pm$ 0.0001 &\cd           &0.0117$\pm$0.0003 &309$_{-12  }^{+12  }$  & 0.12 $\pm$ 0.03\\
010921\bd\bd&0.45 &1.8E-6 $\pm$ 1.6E-7 & \cd                   & \cd               &0.90$\pm$0.30 &0.0014$\pm$0.0015 &89 $_{-13.8}^{+21.8}$  & 3.90 $\pm$ 0.50\\
011211\bd\bd&2.14 &9.2E-8 $\pm$ 9.3E-9 &9.20E-6 $\pm$ 9.5E-7 &0.0044$\pm$ 0.0011 &\cd           &\cd               &59 $_{-8   }^{+8   }$  & \cd            \\
020124\bd\bd&3.20 &6.1E-7 $\pm$ 1.0E-7 &1.14E-5 $\pm$ 1.1E-6 &0.0039$\pm$ 0.0010 &0.08$\pm$0.05 &0.0131$\pm$0.0026 &87 $_{-12  }^{+18  }$  & 0.25 $\pm$ 0.05\\
020405\bd\bd&0.70 &7.4E-6 $\pm$ 3.1E-7 &1.10E-4 $\pm$ 2.1E-6 &0.0060$\pm$ 0.0020 &\cd           &0.0129$\pm$0.0008 &364$_{-90  }^{+90  }$  & 0.45 $\pm$ 0.08\\
020813\bd\bd&1.25 &3.8E-6 $\pm$ 2.6E-7 &1.59E-4 $\pm$ 2.9E-6 &0.0012$\pm$ 0.0003 &0.16$\pm$0.04 &0.0131$\pm$0.0003 &142$_{-13  }^{+14  }$  & 0.82 $\pm$ 0.10\\
020903\bd\bd&0.25 &3.4E-8 $\pm$ 8.8E-9 & \cd                 & \cd               &\cd           &\cd               &2.6$_{-0.8 }^{+1.4 }$  & \cd            \\
021004\bd\bd&2.32 &2.3E-7 $\pm$ 5.5E-8 &3.61E-6 $\pm$ 8.6E-7 &0.0109$\pm$ 0.0027 &0.60$\pm$0.40 &0.0038$\pm$0.0049 &80 $_{-23  }^{+53  }$  & 0.35 $\pm$ 0.15\\
021211\bd\bd&1.01 &2.3E-6 $\pm$ 1.7E-7 & \cd                 & \cd               &0.32$\pm$0.04 &\cd               &46 $_{-6   }^{+ 8  }$  & 0.33 $\pm$ 0.05\\
030115\bd\bd&2.50 &3.2E-7 $\pm$ 5.1E-8 & \cd                 & \cd               &0.40$\pm$0.20 &0.0061$\pm$0.0042 &83 $_{-22  }^{+53  }$  & 1.47 $\pm$ 0.50\\
030226\bd\bd&1.98 &2.6E-7 $\pm$ 4.7E-8 &8.33E-6 $\pm$ 9.8E-7 &0.0034$\pm$ 0.0008 &0.30$\pm$0.30 &0.0058$\pm$0.0047 &97 $_{-17  }^{+27  }$  & 0.70 $\pm$ 0.20\\
030323\bd\bd&3.37 &1.2E-7 $\pm$ 6.0E-8 & \cd                 & \cd               &\cd           &\cd               &44 $_{-26  }^{+90  }$  & 1.00 $\pm$ 0.50\\
030328\bd\bd&1.52 &1.6E-6 $\pm$ 1.1E-7 &6.14E-5 $\pm$ 2.4E-6 &0.0020$\pm$ 0.0005 &0.20$\pm$0.20 &0.0053$\pm$0.0007 &126$_{-13  }^{+14  }$  & \cd            \\
030329\bd\bd&0.17 &2.0E-5 $\pm$ 1.0E-6 &2.31E-4 $\pm$ 2.0E-6 &0.0049$\pm$ 0.0009 &0.14$\pm$0.04 &0.0097$\pm$0.0002 &68 $_{-2.2 }^{+2.3 }$  & 0.66 $\pm$ 0.08\\
030429\bd\bd&2.66 &2.0E-7 $\pm$ 5.4E-8 &1.13E-6 $\pm$ 1.9E-7 &0.0060$\pm$ 0.0029 &\cd           &0.0055$\pm$0.0057 &35 $_{-8   }^{+12  }$  & 0.90 $\pm$ 0.20\\
030528\bd\bd&0.78 &1.6E-7 $\pm$ 3.2E-8 & \cd                 & \cd               &12.5$\pm$0.50 &0.0022$\pm$0.0019 &32 $_{-5.0 }^{+4.7 }$  & 0.77 $\pm$ 0.20\\
040924\bd\bd&0.86 &2.6E-6 $\pm$ 2.8E-7 & \cd                 & \cd               &0.30$\pm$0.04 &\cd               &67 $_{-6   }^{+6   }$  & 0.17 $\pm$ 0.02\\
041006\bd\bd&0.71 &2.5E-6 $\pm$ 1.4E-7 &1.75E-5 $\pm$ 1.8E-6 &0.0012$\pm$ 0.0003 &\cd           &0.0077$\pm$0.0003 &63 $_{-13  }^{+13  }$  & 0.65 $\pm$ 0.16\\
050126\bd\bd&1.29 &1.1E-7 $\pm$ 1.3E-8 & \cd                 & \cd               &2.10$\pm$0.30 &0.0039$\pm$0.0015 &47 $_{-8   }^{+23  }$  & 3.90 $\pm$ 0.80\\
050318\bd\bd&1.44 &5.2E-7 $\pm$ 6.3E-8 &3.46E-6 $\pm$ 3.5E-7 &0.0020$\pm$ 0.0006 &\cd           &0.0071$\pm$0.0009 &47 $_{-8   }^{+15  }$  & 0.38 $\pm$ 0.05\\
050319\bd\bd&3.24 &2.3E-7 $\pm$ 3.6E-8 & \cd                 & \cd               &\cd           &0.0028$\pm$0.0022 &  \cd               & 0.19 $\pm$ 0.04\\
050401\bd\bd&2.90 &2.1E-6 $\pm$ 2.2E-7 & \cd                 & \cd               &0.10$\pm$0.06 &0.0135$\pm$0.0012 &118$_{-18  }^{+ 18 }$  & 0.03 $\pm$ 0.01\\
050406\bd\bd&2.44 &4.2E-8 $\pm$ 1.1E-8 & \cd                 & \cd               &0.64$\pm$0.40 &\cd               &25 $_{-13  }^{+ 35 }$  & 0.50 $\pm$ 0.30\\
050408\bd\bd&1.24 &1.1E-6 $\pm$ 2.1E-7 & \cd                 & \cd               &0.25$\pm$0.10 &\cd               &\cd                    & 0.25 $\pm$ 0.08\\
050416\bd\bd&0.65 &5.3E-7 $\pm$ 8.5E-8 & \cd                 & \cd               &\cd           &\cd               &15 $_{-2.7 }^{+ 2.3}$  & 0.51 $\pm$ 0.30\\
050502\bd\bd&3.79 &4.3E-7 $\pm$ 1.2E-7 & \cd                 & \cd               &0.20$\pm$0.20 &0.0221$\pm$0.0029 &93 $_{-35  }^{+ 55 }$  & 0.40 $\pm$ 0.20\\
050505\bd\bd&4.27 &3.2E-7 $\pm$ 5.4E-8 &6.20E-6 $\pm$ 8.5E-7 &0.0014$\pm$ 0.0007 &\cd           &0.0035$\pm$0.0019 &70 $_{-24  }^{+ 140}$  & 0.40 $\pm$ 0.15\\
050525\bd\bd&0.61 &5.2E-6 $\pm$ 7.2E-8 &2.59E-5 $\pm$ 1.3E-6 &0.0025$\pm$ 0.0010 &0.11$\pm$0.02 &0.0135$\pm$0.0003 &81 $_{-1.4 }^{+ 1.4}$  & 0.32 $\pm$ 0.03\\
050603\bd\bd&2.82 &9.7E-6 $\pm$ 6.0E-7 & \cd                 & \cd               &0.03$\pm$0.03 &0.0163$\pm$0.0015 &344$_{-52  }^{+ 52 }$  & 0.17 $\pm$ 0.02\\
050802\bd\bd&1.71 &5.0E-7 $\pm$ 7.3E-8 & \cd                 & \cd               &\cd           &0.0046$\pm$0.0053 &\cd                    & 0.80 $\pm$ 0.20\\
050820\bd\bd&2.61 &3.3E-7 $\pm$ 5.2E-8 & \cd                 & \cd               &0.70$\pm$0.30 &\cd               &246$_{-40  }^{+ 76 }$  & 2.00 $\pm$ 0.50\\
050824\bd\bd&0.83 &9.3E-8 $\pm$ 3.8E-8 & \cd                 & \cd               &\cd           &\cd               &\cd                    & 11.0 $\pm$ 2.00\\
050904\bd\bd&6.29 &2.5E-7 $\pm$ 3.5E-8 &2.00E-5 $\pm$ 2.0E-6 &0.0097$\pm$ 0.0024 &\cd           &0.0023$\pm$0.0026 &436$_{-90  }^{+ 200}$  & 0.60 $\pm$ 0.20\\
050908\bd\bd&3.35 &9.8E-8 $\pm$ 1.5E-8 & \cd                 & \cd               &\cd           &\cd               &41 $_{-5   }^{+ 9  }$  & 1.50 $\pm$ 0.30\\
050922\bd\bd&2.20 &2.0E-6 $\pm$ 7.3E-8 & \cd                 & \cd               &0.06$\pm$0.02 &0.0033$\pm$0.0006 &198$_{-22  }^{+ 38 }$  & 0.13 $\pm$ 0.02\\
051022\bd\bd&0.80 &1.1E-5 $\pm$ 8.7E-7 &3.40E-4 $\pm$ 1.2E-5 &0.0029$\pm$ 0.0001 &\cd           &0.0122$\pm$0.0004 &510$_{-20  }^{+ 22 }$  & 0.19 $\pm$ 0.04\\
051109\bd\bd&2.35 &7.8E-7 $\pm$ 9.7E-8 & \cd                 & \cd               &\cd           &\cd               &161$_{-35  }^{+ 130}$  & 1.30 $\pm$ 0.40\\
051111\bd\bd&1.55 &3.9E-7 $\pm$ 5.8E-8 & \cd                 & \cd               &1.02$\pm$0.10 &0.0024$\pm$0.0007 &\cd                    & 3.20 $\pm$ 1.00\\
060108\bd\bd&2.03 &1.1E-7 $\pm$ 1.1E-7 & \cd                 & \cd               &\cd           &0.0032$\pm$0.0058 &65 $_{-10  }^{+ 600}$  & 0.40 $\pm$ 0.20\\
\hline
\end{tabular}
\caption{The dataset of GRB observables used in our analysis (from Schaefer
2007).}
\label{tab:grbresults}
\end{table*}

\begin{table*}
\tabcolsep 3pt
\begin{tabular}{|lllllllll|}
\hline
{\bf GRB} & z& $P_{\rm bolo}$ & $S_{\rm bolo}$ & $F_{\rm beam}$ & $\tau_{lag}$ &
       $V$ & $E_{\rm peak}$ &
       $\tau_{RT}$ \\
             &  &[${\rm erg/cm^{2}s}$] & [${\rm erg/cm^{2}}$]& &[sec]& &[keV] &[sec]  \\

\hline
060115\bd\bd&3.53 &1.3E-7 $\pm$ 1.6E-8 & \cd                 & \cd               &\cd           &\cd               &62 $_{-6   }^{+ 19 }$  & 0.40 $\pm$ 0.20\\
060116\bd\bd&6.60 &2.0E-7 $\pm$ 1.1E-7 & \cd                 & \cd               &\cd           &\cd               &139$_{-36  }^{+ 400}$  & 1.30 $\pm$ 0.50\\
060124\bd\bd&2.30 &1.1E-6 $\pm$ 1.2E-7 &3.37E-5 $\pm$ 3.4E-6 &0.0021$\pm$ 0.0002 &0.08$\pm$0.04 &0.0140$\pm$0.0020 &237$_{-51  }^{+ 76 }$  & 0.30 $\pm$ 0.10\\
060206\bd\bd&4.05 &4.4E-7 $\pm$ 1.9E-8 & \cd                 & \cd                 &0.10$\pm$0.10 &0.0025$\pm$0.0016 &75 $_{-12  }^{+ 12 }$  & 1.25 $\pm$ 0.25\\
060210\bd\bd&3.91 &5.5E-7 $\pm$ 2.2E-8 &1.94E-5 $\pm$ 1.2E-6 &0.0005$\pm$ 0.0001 &0.13$\pm$0.08 &0.0019$\pm$0.0004 &149$_{-35  }^{+ 400}$  & 0.50 $\pm$ 0.20\\
060223\bd\bd&4.41 &2.1E-7 $\pm$ 3.7E-8 & \cd                 & \cd               &0.38$\pm$0.10 &0.0075$\pm$0.0033 &71 $_{-10  }^{+ 100}$  & 0.50 $\pm$ 0.10\\
060418\bd\bd&1.49 &1.5E-6 $\pm$ 5.9E-8 & \cd                 & \cd               &0.26$\pm$0.06 &0.0070$\pm$0.0005 &230$_{-[20]}^{+ [20]}$ & 0.32 $\pm$ 0.08\\
060502\bd\bd&1.51 &3.7E-7 $\pm$ 1.6E-7 & \cd                 & \cd               &3.50$\pm$0.50 &0.0010$\pm$0.0017 &156$_{-33  }^{+ 400 }$ & 3.10 $\pm$ 0.30\\
060510\bd\bd&4.90 &1.0E-7 $\pm$ 1.7E-8 & \cd                 & \cd     &\cd    &0.0028$\pm$0.0019 &95 $_{-[30]}^{+ [60]}$ & \cd            \\
060526\bd\bd&3.21 &2.4E-7 $\pm$ 3.3E-8 &1.17E-6 $\pm$ 1.7E-7 &0.0034$\pm$ 0.0014 &0.13$\pm$0.03 &0.0112$\pm$0.0039 &25 $_{-[5] }^{+ [5] }$ & 0.20 $\pm$ 0.05\\
060604\bd\bd&2.68 &9.0E-8 $\pm$ 1.6E-8 & \cd                 & \cd               &5.00$\pm$1.00 &\cd               &40 $_{-[5] }^{+ [5] }$ & 0.60 $\pm$ 0.20\\
060605\bd\bd&3.80 &1.2E-7 $\pm$ 5.5E-8 & \cd                 & \cd               &5.00$\pm$3.00 &\cd               &169$_{-[30]}^{+ [200]}$& 2.00 $\pm$ 0.50\\
060607\bd\bd&3.08 &2.7E-7 $\pm$ 8.1E-8 & \cd                 & \cd               &2.00$\pm$0.50 &0.0059$\pm$0.0014 &120$_{-17}^{+ 190}   $ & 2.00 $\pm$ 0.20\\

\hline
\end{tabular}
\label{tab:grbresults1}
\end{table*}

\begin{figure*}
\mbox{\epsfxsize=12cm \epsffile{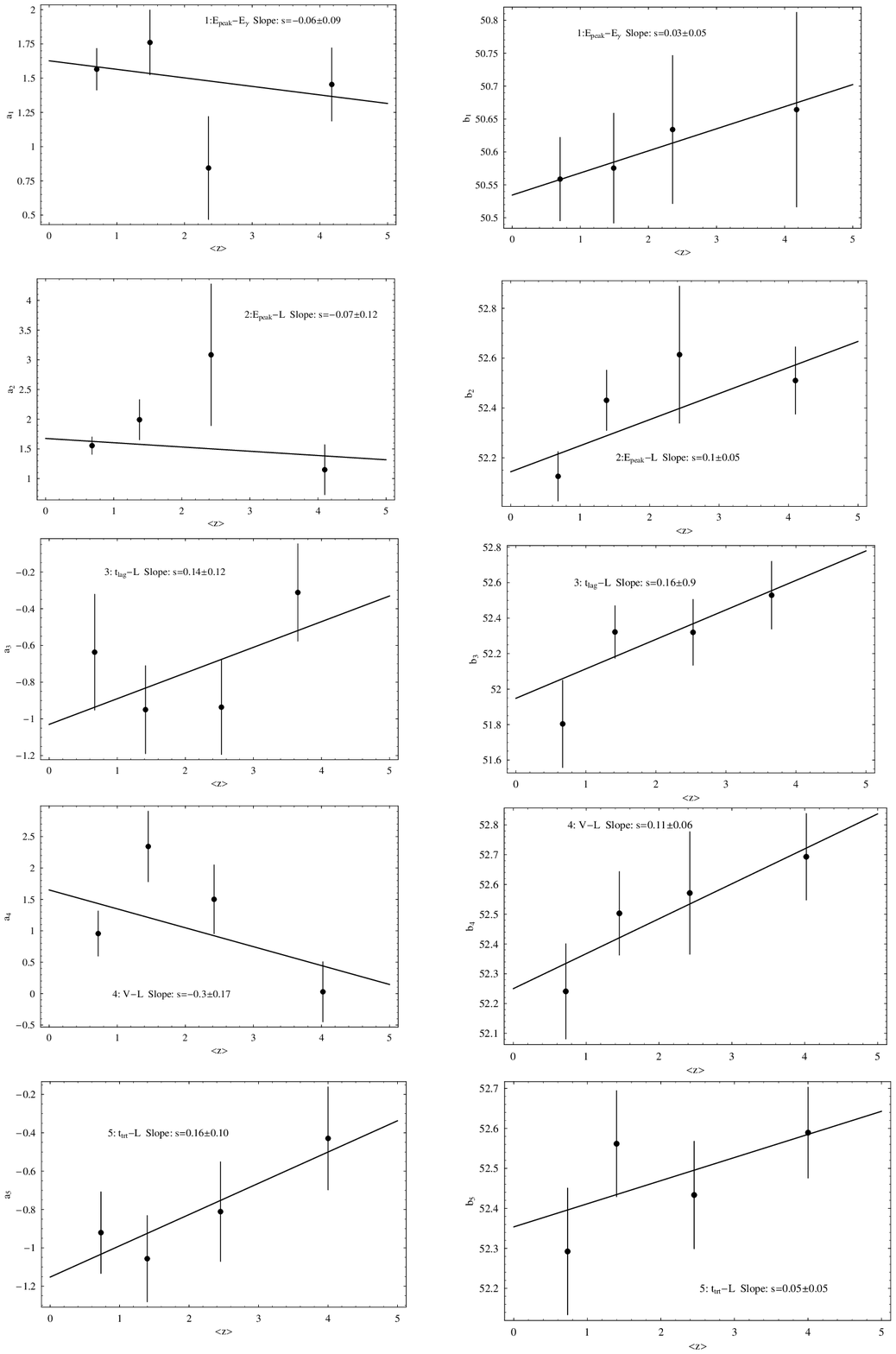}}
\caption{The correlation coefficients $a_j$, $b_j$ ($\omm=0.26$)
obtained in four redshift bins for the five correlation relations
considered. There is no statistically significant indication for
evolution for any of the five correlations considered. \label{fig1}}
\end{figure*}

\indent To explain the calibration procedure in general, we denote
the five luminosity relations by $R_i=B\,Q_i^{a}$ and we take their
logarithms to express them as a linear relation of the form

 \be
 \log{R_i}=\log{B}+a\,\log{Q_i} \Rightarrow  y_i= b+a\,x_i \;
 \label{rab}
\ee
 where $y_i \equiv \log R_i$, $x_i\equiv \log Q_i$ and $b\equiv
\log B$. Notice that $R_i$ depends on the cosmological parameters
[eg. $\omm$) through the luminosity distance (see eqs (\ref{ldl})
and (\ref{egdl})].

In order to find the best fit values of the parameters $a$ and $b$
we use a symmetrized form of the maximum likelihood method
and minimize $\chi^2(a,b,\Omega_{\rm m})$ defined as

\begin{equation}
\label{chi2def}
\chi^2(a,b,\Omega_{\rm m})=
\sum_{i=1}^N\frac{[y_i(\omm) - {\bar y}_i
(a,b)]^2} {\sigma_{y_i}^2 + \sigma_{sys}^2 +a^2 \sigma_{x_i}^2}  \;\;.
\end{equation}
Note that
each correlation point is weighted by its error$^{-1}$ which
means that points with large errors have a negligible contribution to the
$\chi^{2}(a,b,\Omega_{\rm m})$ function. Furthermore,
we normalize $\chi^2$ over the number of points\footnote{Notice that the corresponding errors
assumed to follow a normal distribution, which is the
usual requirement in order
to have a $\chi^2$ distribution.}
and since the number of points is not
too low (see last column of Table 2)
we do not anticipate
the number of points by itself to introduce a further significant bias.
The quantities appearing in eq. (\ref{chi2def}) are defined as
follows:
\begin{itemize}\item $y_i(\omm)$ corresponds to
either $\log L$ or $\log E_\gamma$ as
obtained from Table 1 and eqs. (\ref{ldl})-(\ref{egdl}) and depends
on the cosmological parameter $\omm$ through the luminosity distance
$d_L (\omm,z_i)$. \item ${\bar y}_i (a,b)$ is the predicted value of
$y_i\equiv \log R_i$ on the basis of the linear logarithmic relation
(\ref{rab}). \item Using error propagation we find $\sigma_{y_i}$
from the errors of Table 1 as
\be
 \sigma_{y_i}= \left[\frac{dy(R_i)}{dR_i}\right] \sigma_{R_i} =
\frac{1}{\ln{10}}\frac{\sigma_{P_{\rm bolo} } }{P_{\rm bolo}}
 \label{sldl}
 \ee
 when $y$ corresponds to $\log L$ and
\be
\sigma_{y_i} =\frac{1}{\ln{10}}
\sqrt{\left( \frac{\sigma_{\rm S_{bolo}}}{S_{\rm bolo}} \right)^{2}+
\left( \frac{\sigma_{\rm F_{beam}}}{F_{\rm beam}} \right)^{2} }
\label{sdegama} \ee
when $y$ corresponds to $\log
E_\gamma$. In the case of
asymmetric errors we utilize the simplest approach
of symmetrizing the errors by taking their mean value on
the two sides.
Similarly, we find $\sigma_{x_i}$ as
\be
 \sigma_{x_i}= \left[\frac{dx(Q_i)}{dQ_i}\right] \sigma_{Q_i} =
\frac{1}{\ln{10}}\frac{\sigma_{Q_{i} } }{Q_{i}}\; .
 \label{sx}
 \ee
\item $\sigma_{sys}$ is an assumed additional source of intrinsic
scatter determined for each correlation relation.
Note that we do not treat $\sigma_{sys}$ as a free parameter
(we do not minimize with respect to it)
but we fix it only at the end of the minimization by demanding
$N\chi^{2}/dof\simeq \chi^2=O(1)$.
Thus we treat $\sigma_{sys}$ not as a
physics related parameter but as an unknown source of scatter which
is required to make the quality of fit acceptable. A similar approach has
been used by the SNLS collaboration (Astier et al. 2006) in cosmological
fits of type Ia supernova data. This approach however should be generally used
with care (Vishwakarma 2007) since it may hide a low quality of distance
indicators or it may introduce a good quality of fit for a model, by brute force.
\end{itemize}
Equation (\ref{chi2def}) has the advantage of being symmetric with
respect to errors in both the $x_{i}$ and $y_{i}$ variables
(see eg Amati et al. 2008 and references therein). 
It is interesting to mention here that the statistical results
depend on where we put the $\sigma_{sys}$ parameter (either in
$x$, $y$-axis or both). For example $\sigma_{sys}$ could have been included along with the $x$-axis error as
\begin{equation}
\label{chi2def1}
\chi^2(a,b,\Omega_{\rm m})=
\sum_{i=1}^N\frac{[y_i(\omm) - {\bar y}_i
(a,b)]^2} {\sigma_{y_i}^2 + \sigma_{sys}^2 +a^2 (\sigma_{x_i}^2+\sigma_{sys}^2)}  \;\;.
\end{equation} 
leading to full symmetry between $x$ and $y$. We have verified 
numerically that the use of eq. (\ref{chi2def1}) [proposed
by N. Butler private communication] leads to values 
of best fit parameters $a$ and $b$ that depend very sensitively on 
the value of $\sigma_{sys}$. We consider this to be an undesirable 
feature and thus we have chosen to use the more robust form of (\ref{chi2def}).
Schaefer (2007) on the other hand, uses a linear regression procedure
by putting $\sigma_{sys}$ in both $x$ and $y$.
The novelty of our statistical analysis is that we treat
the problem with a maximum likelihood which includes the data errors and
thus the corresponding results are less dependent on the assumption
related with the value of the $\sigma_{sys}$ parameter.
\begin{figure*}
\mbox{\epsfxsize=12cm \epsffile{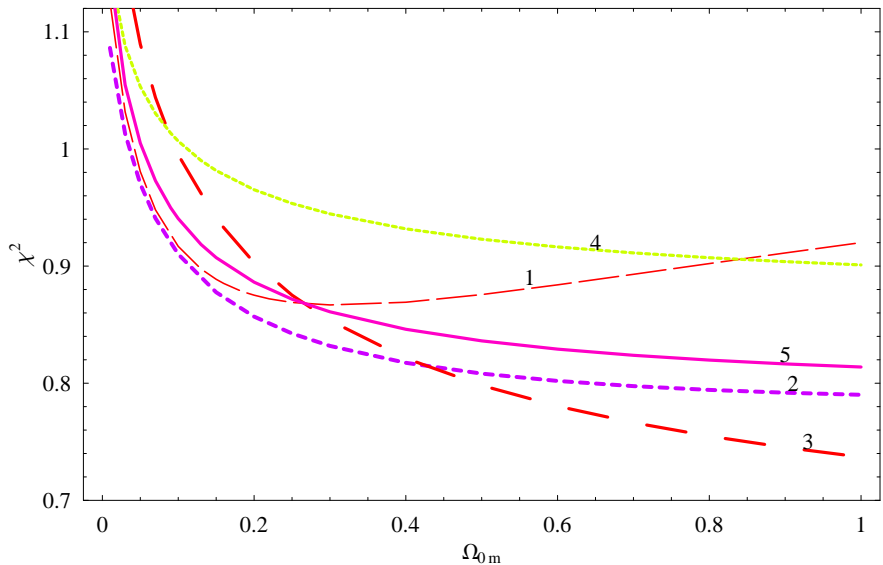}}
\caption{The variation of $\chi^2$ as a function of the cosmological
parameter $\omm$ in a flat $\Lambda$CDM cosmology. Notice that only the
highest quality correlation relations $1$ favors (mildly) an
accelerating universe ($\omm\simeq 0.26$). Minimization with respect
to $a$, $b$ was performed for each value of $\omm$.}
\end{figure*}
A symmetrized method may be warranted because 
the two variables $R_i$, $Q_i$ are not directly causative
[for example, it is possible
that the scatter may be dominated by hidden variables, or
variables not directly measured or treated, such as
the bulk Lorentz factor $\Gamma_{jet}$
(Schaefer et al. 2003, Amati et al. 2002)].
The bisector method of two ordinary least squares
(Isobe et al. 1990) used in previous studies (Schaefer 2007)
is an alternative symmetrized approach (for other similar approaches
see Kelly 2007). That approach however
ignores the measurement uncertainties during the fit which are taken
into account in our symmetrized maximum likelihood method. The
advantage of expressing $\chi^2(a,b,\omm)$ in terms of both the
calibration parameters $a$, $b$ and the cosmological parameter
$\omm$ is that it allows either fixing $\omm$ and calibrating $a$,
$b$ to be used for constraining other cosmological models
(Schaefer 2007) or minimizing simultaneously with respect to
all three parameters (Li et al. 2008b; Qi, Wang \& Lu 2008). Even though the
former approach may be useful as a consistency test it suffers from
a circularity problem since it assumes a particular
cosmological model to make the calibration and thus introduces a
bias against alternative models. The later approach bypasses this
circularity problem at the expense of increasing the uncertainties
in the parameter determination since there are now three parameters
to be fit instead of just two. Thus, in the next section we use the
former approach (fixing $\omm$) to investigate the possible redshift
dependence of the calibration parameters $a$, $b$ but we use the
later approach (variable $\omm$) when testing for the model
dependence of the calibration.

\section{Results of Calibration Tests}
\subsection{Redshift Dependence of Calibration}
In order to test if the correlation relations discussed in the
previous section vary with redshift we separate the GRB samples in
each case into four groups corresponding to the following redshift
bins: $z\in [0,1]$, $z\in [1,2]$, $z\in [2,3]$ and $z\in [3,7]$. The
number of GRBs corresponding to each correlation relation and each
redshift bin is shown in Table 2.

\begin{table}
{\small
\caption[]{Number of GRBs in each redshift bin.}
\tabcolsep 2pt
\begin{tabular}{ccccccc}
\hline
   Correl. Type & $z\in [0,1]$  & $z\in [1,2]$ & $z\in [2,3]$ & $z\in
   [3,7]$ & Total \\ \hline
$\epkk-E_\gamma$&10 & 8& 4 & 5 & 27\\
$\epkk-L$&18 &15 &14 &17 & 64 \\
$\tlag-L$&7 &13 & 9 & 9 & 38\\
$V-L$&14 &15 &9 &13 & 51\\
$\trt-L$&17&15 &13 &17 & 62 \\ \hline
\end{tabular}
}
\end{table}

We now perform a maximum likelihood fit in each group after setting
$\omm =0.26$
obtained from the five years WMAP data
(Komatsu et al. 2008),
and determine the best fit calibration parameters $a$,
$b$ with $1\sigma$ errors and the quality of fit expressed through
the minimum $\chi^2$ (per degree of freedom). We use a different
value for the intrinsic dispersion $\sigma_{sys}$ for each type of
correlation obtained by demanding a value of $\chi^2$ of order 1.
Thus we transfer the information about the quality from $\chi^2$ to
$\sigma_{sys}$ (the smaller the required $\sigma_{sys}$ the better
the fit). These results are shown in Table 3. We have verified
that the main features of our results do not depend on the choice of
$\omm$ in the range $\omm \in [0.2,0.3]$. Despite of the fact that
we use the maximum likelihood method instead of linear regression
used in Schaefer (2007) our results for $a_j$, $b_j$ in
the full redshift range (last column of Table 3) are consistent at
the $1\sigma$ level with those of Schaefer (2007).

\begin{table*}
 \caption{\label{tablebs2007} Best fit parameters in each redshift bin}
 \begin{tabular}{c c c c c c c}
   Correlation Type & $z\in [0,1]$  & $z\in [1,2]$ & $z\in [2,3]$ & $z\in [3,7]$ & Total \\
\hline
\\
$\epkk-E_\gamma$&$b=50.56\pm 0.06$&$b=50.57\pm0.08$ & $b=50.63\pm 0.11$& $b=50.66\pm 0.15$& $b=50.60\pm 0.04$ \\
 ($\sigma_{sys}=0.15$) &$a=1.56\pm 0.15$ &$a=1.76\pm 0.23$ & $a=0.84\pm 0.37$ & $a=1.45\pm 0.26$  & $a=1.56\pm 0.11$  \\
  &$\chi^2=1.03$ &$\chi^2=0.65$& $\chi^2=0.51$ & $\chi^2=0.34$ & $\chi^2=0.86$  \\
 \hline
\\
$\epkk-L$&$b=52.12\pm 0.10$&$b=52.43\pm 0.12$ & $b=52.61\pm 0.27$& $b=52.51\pm 0.13$& $b=52.33\pm 0.06$ \\
 ($\sigma_{sys}=0.30$) &$a=1.55\pm 0.14$ &$a=1.99\pm 0.33$ & $a=3.08\pm 1.19$ & $a=1.14\pm 0.41$  & $a=1.75 \pm 0.11$  \\
  &$\chi^2=1.11$ &$\chi^2=0.89$ & $\chi^2=0.40$ & $\chi^2=0.20$ & $\chi^2=0.84$  \\
\hline
\\
$\tlag-L$&$b=51.80\pm 0.24$&$b=52.32\pm 0.14$ & $b=52.32\pm 0.18$& $b=52.52\pm 0.19$& $b=52.24\pm 0.09$ \\
 ($\sigma_{sys}=0.39$) &$a=-0.63\pm 0.31$ &$a=-0.94\pm0.23$ & $a=-0.93\pm 0.25$ & $a=-0.31\pm 0.26$  & $a=-0.87\pm 0.12$  \\
  &$\chi^2=0.62$ &$\chi^2=0.70$ & $\chi^2=0.63$ & $\chi^2=0.13$  & $\chi^2=0.87$  \\
\hline
\\
$V-L$&$b=52.24\pm 0.16$&$b=52.50\pm 0.14$ & $b=52.57\pm 0.20$& $b=52.69\pm 0.14$& $b=52.45\pm 0.07$ \\
 ($\sigma_{sys}=0.47$) &$a=0.96\pm 0.36$ &$a=2.34\pm 0.56$ & $a=1.50\pm 0.54$ & $a=0.03\pm 0.47$  & $a=1.30\pm 0.21$  \\
  &$\chi^2=1.22$ &$\chi^2=0.57$ & $\chi^2=0.58$ & $\chi^2=0.32$  & $\chi^2=0.95$  \\
\hline
\\
$\trt-L$&$b=52.29\pm 0.16$&$b=52.56\pm 0.13$ & $b=52.43\pm 0.13$& $b=52.59\pm 0.11$& $b=52.46\pm 0.06$ \\
 ($\sigma_{sys}=0.48$) &$a=-0.92\pm 0.21$ &$a=-1.05\pm 0.22$ & $a=-0.81\pm 0.26$ & $a=-0.43\pm 0.26$  & $a=-0.92\pm 0.11$  \\
  &$\chi^2=0.52$ &$\chi^2=0.27$ & $\chi^2=0.56$ & $\chi^2=0.19$  & $\chi^2=0.87$  \\
\hline
\\
\end{tabular}
\footnotetext[1]{The column `Total' corresponds to the full set of
data.}
 \end{table*}

The corresponding plots of the calibration parameters $a_j$, $b_j$
vs $<z>$ (average redshift in each redshift bin) are shown in Fig. 1
for all five correlation relations.
The slope of each $a_j(<z>)$, $b_j(<z>)$ with $1\sigma$ errors is also shown in Fig. 1. Notice that 
all slopes are consistent with zero at the $2\sigma$ level.
Thus,
there is no statistically significant evidence for evolution with
redshift of the correlation relations. This result is to be
contrasted with the result of Li (2007a, 2007b) where the Amati et al. (2002)
relation $E_{peak}-E_{\gamma-iso}$ was found to
have statistically significant correlation with redshift for both
parameters $a$ and $b$. However the latter results are under dispute by
the recent paper of Ghirlanda et al. (2008). Here we use an improved variant of this
relation namely $E_{peak}-E_{\gamma}$ (Ghirlanda et al. 2004) and a
somewhat different GRB dataset (Schaefer 2007) and we find no
statistically significant evidence for correlation of $a$, $b$ with
redshift. It is also interesting to use the results of Table 3 in order to
compare the five correlation relations with respect to their quality
of linear fit. Since the value of the intrinsic dispersion
$\sigma_{sys}$ has been adjusted for each correlation so that
$\chi^2\simeq 1$, the comparison can not be made by simply comparing the
values of $\chi^2$ (per degree of freedom) in each case. Instead we
compare the required value of $\sigma_{sys}$. Smaller $\sigma_{sys}$
corresponds to better quality of fit. According to this test, the
correlation relations are ranked according to their quality as
follows: \begin{enumerate} \item $\epkk-E_\gamma$
($\sigma_{sys}=0.15$), \item $\epkk-L$ ($\sigma_{sys}=0.30$),
\item $\tlag-L$ ($\sigma_{sys}=0.39$), \item $V-L$ ($\sigma_{sys}=0.47$),
\item $\trt-L$ ($\sigma_{sys}=0.48$). \end{enumerate}
Clearly, the correlation relation $\epkk-E_\gamma$ provides
significantly better fit compared to the other four relations.

\begin{figure}
\mbox{\epsfxsize=8.5cm \epsffile{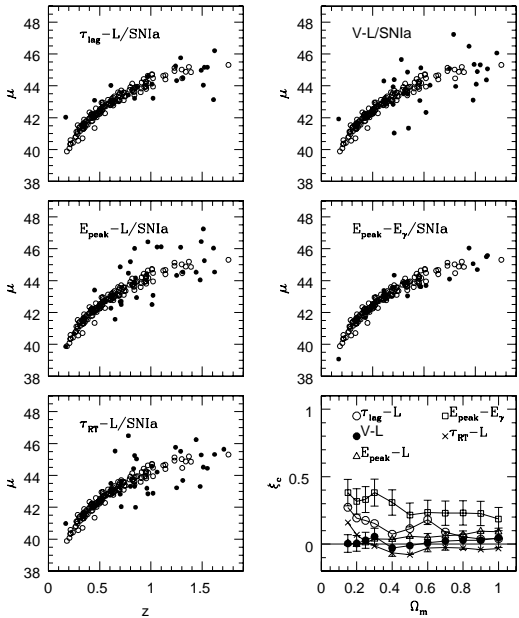}}
\caption{
Comparison of the distance modulus as a function of
redshift, estimated by the GRB observations (solid points) and the a
recent SnIa dataset of Davis et al. (open points). {\it Last
Panel:} The average cross-correlation function between the GRBs
and SnIa as a function of the density parameter. The different
points correspond to different luminosity indicators for the current
GRB data and thus to different values of the fitted model
parameters. Note that we don't plot all the error-bars in order to
avoid confusion.}
\end{figure}

\subsection{Model Dependence of Calibration}
In order to investigate the cosmological model dependence of the
calibration parameters $a_j$, $b_j$ we now allow the cosmological
parameter $\omm$ to vary in the $\chi^2(a,b,\omm)$ of eq.
(\ref{chi2def}). In particular we minimize $\chi^2(a,b,\omm)$ with
respect to $a_j$, $b_j$ for various $\omm \in [0.2,0.3]$. We show
$\chi^2_{min}(\omm)$ in Fig. 2 for all five correlation relations.
As shown in Fig. 2 the best correlation relation ($\epkk-E_\gamma$)
develops a minimum with respect to $\omm$ at
$\omm=0.30^{+1.60}_{-0.25}$ (the $1\sigma$ error is obtained by
demanding $\chi^2 < \chi^2_{min}+ 3.53/N$ corresponding to maximum
likelihood minimization with three parameters). On the other hand,
the $\chi^2$ for other four correlation relations are monotonic with
respect to $\omm$ and seem to favor a flat matter dominated
universe. We attribute this behavior to the large intrinsic scatter
inherent in these correlation relations. The fact that these
relations favor large $\omm$ may explain the fact that the best fit
$\omm$ obtained by Schaefer (2007) using a combination
of all five correlation was somewhat larger ($\omm \simeq 0.39$)
than the corresponding values obtained using
other more robust cosmological tests. We thus argue that the mixing
of the best quality correlation with the other four correlations
plagued with large scatter may lead to misleading results and
perhaps should be avoided. To this end, it is interesting to mention
that systematic effects introduced by the flux limits
of the GRB surveys (Lloyd et al. 2000) could potentially affect
the above statistical results.
In order to check such a  possibility, in the next section
we perform a direct comparison between the GRBs with the SnIa data.

\section{GRBs versus SnIa}
An alternative model independent test of the GRBs as standard candles
is their direct comparison with the best cosmological standard
candles available namely SnIa. Thus we perform a cross-correlation
analysis in the distance modulus space in the redshift range $z\in
[0.17,1.75]$. The cross-correlation between two samples (in our case
GRB and SnIa) is typically defined by the following
estimator (Peebles 1973; Efstathiou et al. 1991):
\be \xi(\mu)=f\frac{N_{GG}}{N_{GS}}-1,
\ee where $N_{GG}$ and $N_{GS}$ is the number of GRB-GRB and
GRB-SnIa pairs respectively in the
interval $[\mu, \mu + \delta\mu]$.
For a $\xi(\mu)=0$ there is not any correlation between the
two populations.
Having known (from the
$\chi^{2}-$analysis) the calibrations for every $L$ or $E_{\gamma}$,
we can also derive [see eqs (\ref{ldl}) and (\ref{egdl})] the
corresponding distance modulus $\mu(z)=5{\rm log}d_{L}(z)+25$. Note
that $\delta\mu$ is set at 0.75 which corresponds to
10 number of bins. The robustness of our results was tested
using different bins (spanning from 7 to 15) and we found
very similar statistical results. In the above relation $f$ is the
normalization factor $f = 2 N_S /(N_G-1)$ where $N_G$ and $N_S$ are
the total number of GRB and SnIa entries respectively. The
uncertainty in $\xi(\mu)$ is estimated as
$\sigma_{\xi}=\sqrt{(1+\xi(\mu))/N_{GS}}$ (Peebles 1973).
To this end, we estimate the average
cross-correlation function, which is given by: \be
\xi_{c}=N^{-1}_{b} \sum_{i=1}^{N_{b}} \xi^{i}(\mu)\ee
where $N_{b}$ is the number of bins used ($N_{b}=10$).

In this paper we utilize the sample of 192 standard candles
(supernovae) of Davis et al. (2007) in the $0.016\le
z\le 1.755$. Due to the fact that the GRB observations extend beyond
$z\ge 0.17$, we apply our statistical analysis to the following
redshift interval $0.170\le z\le 1.755$. In this case, the SnIa
subsample contains 146 entries, $\epkk-E_\gamma$ contains 17
entries, $\epkk-L$ contains 32 entries, $\tlag-L$ contains 19
entries, $V-L$ contains 28 entries and $\trt-L$ contains 31 entries.

In Fig.3 we compare the estimated GRB distance moduli (solid points)
with those derived by the SnIa data (open points) as a function of
redshift. Performing, a standard Kolmogorov-Smirnov (KS)
test to the distance moduli we find that the probability of consistency
between GRBs and SnIa using the $E_{peak}-E_\gamma$ relation
is ${\cal P}_{KS}\sim 0.4$. Clearly,
there is a strong indication that the
$E_{peak}-E_\gamma$ relation traces well the SnIa regime, a fact
corroborated also by the cross-correlation test between GRBs and
SnIa, which gives an average cross-correlation function of
$0.38\pm 0.09$ for $\Omega_{\rm m}=0.30$ or a  $\sim 4\sigma$
correlation signal. Notice that the cross-correlation analysis was
performed for each value of $\Omega_{\rm m}$. Doing so, in the last
panel of Fig.3 we present, for each GRB subsample, the average
cross-correlation function $\xi_{c}$ as a function of $\Omega_{\rm
m}$. It becomes clear that the best GBR tracer of the SnIa regime is
the $E_{peak}-E_\gamma$ relation. Interestingly, the cross
correlation function peaks at $\Omega_{\rm m}\simeq 0.15$ and
$\Omega_{\rm m}\simeq 0.30$ respectively. However, due to small
number statistics the $E_{peak}-E_\gamma$ test (17 entries) provides
much weaker constraints on $\omm$ than current SnIa data. On the
other hand, the $\tau_{lag}-L$ gives a relatively good correlation
signal up to $\Omega_{\rm m}\le 0.3$ although, the amplitude of the
cross-correlation function decreases rapidly as a function of
$\Omega_{\rm m}$ than the $E_{peak}-E_\gamma$ case. Finally, the
$\tau_{RT}-L$, $E_{peak}-L$ and $V-L$ relations seem to fluctuate
around zero (in agreement with the $\chi^{2}$ minimization
results of Fig. 2).

\section{Conclusions}
We have investigated the robustness of the current GRB data
calibrated as luminosity indicators with respect to two sources of
biases: evolution of the calibration with redshift and dependence of
the calibration on the assumed cosmological model. We have 
found no statistically significant evidence for evolution of the
calibration parameters $a$, $b$ with redshift for any of the
luminosity indicator correlations considered.
However, our statistical results do not exclude the possibility
of correlation to be discovered in the future based on better GRB data.
We have also found
that the correlation $\epkk-E_\gamma$ has two important advantages
over the other four correlations considered:
\begin{itemize}\item Its intrinsic scatter $\sigma_{sys}$ is less
than half of the corresponding scatter of the other four
correlations \item It can pick up the accelerating expansion of the
universe in a model independent way [it has a clear global minimum
of $\chi^2(a,b,\omm)$ at $\omm \simeq 0.3$]\item It traces
relatively well the SnIa regime.
\end{itemize}
However, even the best GRB luminosity indicator correlation is
currently not competitive with other cosmological probes of the
acceleration expansion since the cosmological parameter $1\sigma$
errors ($\omm=0.30^{+1.60}_{-0.25}$) are more than an order of magnitude
larger than the corresponding errors obtained eg using SnIa standard
candles and other geometrical probes ($\omm =
0.267^{+0.028}_{-0.018}$). Therefore, even though the GRB data are
currently not competitive with other cosmological probes of the
accelerating expansion of the universe this may well change in the
future if the $\epkk-E_\gamma$ GRB dataset expands well beyond its
current status consisting of only 27 datapoints.

\section*{Acknowledgements} We thank S. Nesseris for useful discussions.
We also thank the referee N. Butler for his very detailed report, useful comments and
suggestions.
This work was supported by the European Research and Training
Network MRTPN-CT-2006 035863-1 (UniverseNet).
{\bf Numerical Analysis:} The mathematica files with the numerical
analysis of this study may be found at
http://leandros.physics.uoi.gr/grb/grb.htm or may be sent by e-mail
upon request.


{\small
\begin{thebibliography}{}
\bibitem[]{}Amati, L., et al., 2002, ApJ, 390, 81
\bibitem[]{}Amati, L., et al., 2008, arXiv:0805.0377, MNRAS, submitted
\bibitem[]{}Astier, P. {\it et al.}, 2006, A\&A,   447, 31
\bibitem[]{}Bertolami, O., \&, Tavares Silva, P., 2006, MNRAS, 365, 1149
\bibitem[]{}Bertschinger, E., 2006, ApJ, 648, 797
\bibitem[]{}Boisseau, B, Esposito-Farese, G.,
Polarski D., \&, Starobinsky, A. A., 2000, Phys. Rev. Lett. 85, 2236
\bibitem[]{}Butler, N. R., Kocevski, D., Bloom, J. S., \&,
Curtis, J. L, 2007, ApJ, 671, 656
\bibitem[]{}Butler, N. R., Kocevski, D., \&, Bloom, J. S., 2008,
submitted to ApJ,  (arXiv0802.3396)
\bibitem[]{}Caldwell, R., Cooray, A., \&, Melchiorri, A., 2007,
Phys. Rev. D, 76, 023507
\bibitem[]{}Dai, Z. G.,  Liang, E. W., \&, Xu, D., 2004, ApJ, 612, L101
\bibitem[]{}Davis, T. M., et al., 2007, ApJ, 666, 716
\bibitem[]{}Demianski, M., Piedipalumbo, E., Rubano, C., \&, Tortora, C., 2006
A\&A, 454, 55
\bibitem[]{}Di Girolamo, T., Catena, R., Vietri, M., \&, Di Sciascio, G., 2005,
  JCAP, 0504, 008
\bibitem[]{}Efstathiou, G., Bernstein, G., Katz, N., Tyson, J. A., \&,
  Guhathakurta, P., 1991, ApJ, 380, L47
\bibitem[]{}Fenimore, E. E., \&, Ramirez-Ruiz, E.,
2000, (arXiv:astro-ph/0004176)
\bibitem[]{}Firmani, C., Ghisellini, G., Avila-Reese, V., \&, Ghirlanda, G.,
2006, MNRAS, 370, 185
\bibitem[]{}Frail, D. A., et al., 2001, ApJ, 562, L55
\bibitem[]{}Ghirlanda, G., Ghisellini, G., \&, Lazzati, D., 2004, ApJ,
  616, 331
\bibitem[]{}Ghirlanda, G., Ghisellini, G., \&, Firmani, C., 2006,
New J. Phys., 8, 123
\bibitem[]{}Ghirlanda, G., Nava, L., Ghisellini, G., Firmani, C., \&,
Cabrera, J. I., 2008, MNRAS, 387, 319
\bibitem[]{}Heavens, A. F., Kitching, T. D., \&, Verde, L.,
2007, MNRAS, 380, 1029
\bibitem[]{}Hooper, D., \&, Dodelson, S., 2007,
Astropart. Phys.,  27, 113
\bibitem[]{}Isobe, T., Feigelson, E. D., Akritas, M. G., \&, Babu,
G. J., 1990, ApJ, 364, 104
\bibitem[]{}Jain B., \&, Zhang, P., 2007, (arXiv:0709.2375)
\bibitem[]{}Kawai, N., et al., 2006, Nature, 440, 184
\bibitem[]{}Kelly, B. C., 2007, ApJ, 665, 1489
\bibitem[]{}Komatsu, E., et al., 2008, ApJS, submitted, (arXiv:0803.0547)
\bibitem[]{}Lazkoz, R., Nesseris, S., \&, Perivolaropoulos, L., 2007,
(arXiv:0712.1232)
\bibitem[]{} Li, L-X., 2007a, MNRAS, 379, L55
\bibitem[]{}Li, L-X, 2007b, To appear in the proceedings of
'The Next Decade of GRB afterglows', Amsterdam
(arXiv:0705.4401)
\bibitem[]{}Li, H., Su, M., Fan, Z., Dai, Z., \&, Zhang, X., 2008a,
Phys. Lett. B, 658, 95
\bibitem[]{}
Li, H., Xia, Jun-Qing, Liu, J., Zhao, Gong-Bo, Fan, Zu-Hui,\&, Zhang, X.
2008b, ApJ, 680, 92
\bibitem[]{}Lloyd, N. M., Petrosian, V., Mallozzi, R. S., 2000, ApJ, 534, 227
\bibitem[]{}Mosquera Cuesta, H. J., Dumet, H. M.,\&, Furlanetto, C., 2007, (arXiv:0708.1355)
\bibitem[]{}Nesseris, S., \&, Perivolaropoulos, L., 2006,
  Phys. Rev. D, 73, 103511
\bibitem[]{}Nesseris, S., \&, Perivolaropoulos, L., 2007,
  Phys. Rev. D, 75, 023517
\bibitem[]{}Nesseris, S., \&, Perivolaropoulos, L., 2008,
  Phys. Rev. D, 77, 023504
\bibitem[]{}Norris, J. P., Marani, G. F., \&, Bonnell, J. T.,
  2000,ApJ, 534, 248
\bibitem[]{}Padmanabhan, T., 2003, Phys. Rept., 380, 235
\bibitem[]{}Peebles, P. J. E., 1973, ApJ, 185, 413
\bibitem[]{}Perivolaropoulos, L, 2005, JCAP, 0510, 001
\bibitem[]{}Perlmutter, S., et al., 1999, ApJ, 517, 565
\bibitem[]{}Percival, W. J., Cole, S., Eisenstein, D. J., Nichol, R. C.,
Peacock, J. A., Pope, A. C., \&, Szalay, A. S., 2007, MNRAS, 381, 1053
\bibitem[]{}Piran, T., 2004, Rev. Mod. Phys., 76, 1143
\bibitem[]{}Qi, S., Wang, F. Y.,\&, Lu, T., 2008, A\&A, 483, 49
\bibitem[]{}Rhoads, J. E., 1999, ApJ, 525, 737
\bibitem[]{}Reichart, D. E., Lamb, D. Q., Fenimore, E. E.,
Ramirez-Ruiz, E., Cline, T. L., Hurley, K., 2001, ApJ, 552, 57
\bibitem[]{}Riess, A. G., et al., 1998, AJ, 116, 1009
\bibitem[]{}Sari, R., Piran, T., \&, Halpern, J. P., 1999, ApJ, 519, L17
\bibitem[]{}Schaefer, B. E., 2003, ApJ, 583, L67
\bibitem[]{}Schaefer, B. E., 2007, ApJ, 660,16
\bibitem[]{}Tegmark, M., et al., 2006, Phys. Rev. D., 74, 123507
\bibitem[]{}Vishwakarma, R.G., 2007, Int. J. Mod. Phys.  D 16, 1641.
  [arXiv:astro-ph/0511628].

\bibitem[]{}Wang, F. Y., \&, Dai, Z. G., 2006, MNRAS, 368,371
\bibitem[]{}Wang, S., Hui, L., May, M., \&, Haiman, Z., 2007
Phys. Rev. D, 76, 063503
\bibitem[]{}Zhang, Z. B.\&,  Xie, G. Z., 2007, (arXiv:0711.1411)
\bibitem[]{}Zhang, B., \&,  Meszaros, P., 2004,
Int. J. Mod. Phys. A, 19, 2385

\end{thebibliography}
}
\end{document}